\documentstyle[11pt,aaspp4,psfig]{article}
\def\etal{{\rm et~al.\ }}

\begin{document}

\title{Gamma-ray limits on Galactic $^{60}$Fe nucleosynthesis and \\
         implications on the Origin of the $^{26}$Al emission}

\author{Juan E. Naya\altaffilmark{1}, Scott D. Barthelmy\altaffilmark{1}, 
Lyle M. Bartlett\altaffilmark{2}, Neil Gehrels, Ann Parsons, Bonnard J. 
Teegarden and Jack Tueller}

\affil{NASA/Goddard Space Flight Center, Greenbelt, MD 20771, USA}

\author{Marvin Leventhal}

\affil{Dept. of Astronomy, University of Maryland, College Park, MD 20742-2421, USA}

\altaffiltext{1}{Universities Space Research Association, 7501 Forbes 
Blvd.  \#206, Seabrook, MD 20706-2253, USA} 

\altaffiltext{2}{NAS/NRC Resident Research Associate, Code 718, 
NASA/GSFC, Greenbelt, MD 20771, USA}

\begin{abstract}

The Gamma Ray Imaging Spectrometer (GRIS) recently observed the 
gamma-ray emission from the Galactic center region.  We have detected 
the 1809 keV Galactic $^{26}$Al emission at a significance level of 
6.8$\sigma$ but have found no evidence for emission at 1173 keV and 
1332 keV, expected from the decay chain of the nucleosynthetic 
$^{60}$Fe.  The isotopic abundances and fluxes are derived for 
different source distribution models.  The resulting abundances are 
between $2.6\pm0.4$ and $4.5\pm0.7$ $\, M_\odot$ for $^{26}$Al and a 
2$\sigma$ upper limit for $^{60}$Fe between 1.7 and 3.1 $\, M_\odot$.  
The measured $^{26}$Al emission flux is significantly higher than that 
derived from the CGRO/COMPTEL 1.8 MeV sky map.  This suggests that a 
fraction of the $^{26}$Al emission may come from extended sources with 
a low surface brightness that are invisible to COMPTEL. We obtain a 
$^{60}$Fe to $^{26}$Al flux ratio 2$\sigma$ upper limit of 0.14, which 
is slightly lower than the 0.16 predicted from current nucleosynthesis 
models assuming that SNII are the major contributors to the galactic 
$^{26}$Al.  Since the uncertainties in the predicted fluxes are large 
(up to a factor of 2), our measurement is still compatible with the 
theoretical expectations.

\end{abstract}


\section{Introduction}

Stellar nucleosynthesis produces a number of radioisotopes which are 
potentially observable through their gamma-ray emission.  
Radioisotopes with decay times that are long compared to the intervals 
between the events that eject them establish steady state abundances 
in the interstellar medium.  Among these radioisotopes, the species 
$^{26}$Al ($t_{1/2}=7.2\cdot10^{5}$ y) and $^{60}$Fe ($t_{1/2}=1.5 
\cdot10^{6}$ y) are of particular interest since their decay times are 
short compared to Galactic rotation.  The abundances of these species 
in the interstellar medium therefore serves as an important tracer of 
the stellar population responsible for their synthesis.  The existence 
of Galactic 1809 keV line radiation was well established after the 
detections by the HEAO-C and SMM spacecraft and was subsequently 
confirmed by different balloon-borne instruments (see Prantzos \& 
Diehl 1996 for a review).  In spite of the enormous progress in the 
understanding of the 1809 keV emission in the last few years, thanks 
to the sky maps obtained by COMPTEL, the main contributor to this 
emission is still an issue of discussion.  Theory predicts that 
$^{26}$Al is released into the interstellar medium by nova and 
supernova explosions, from winds of massive stars in the Wolf-Rayet 
phase, and from less-massive stars in the very late stages of their 
evolution (in the Asymptotic Giant Branch phase).  However, 
uncertainties in the models do not allow the contributions of each 
source to be precisely predicted (\cite{pra96}).  An important clue to 
this puzzle would be provided by the detection of the $^{60}$Fe 
emission.  The isotope $^{60}$Fe ($t_{1/2}=1.5\cdot10^{6}$ y) decays 
to $^{60}$Co ($t_{1/2}=5.3 $ y) which then decays to $^{60}$Ni, 
simultaneously emitting two gamma-ray photons of energies 1173 and 
1332 keV. Models predict that $^{60}$Fe is released into the 
interstellar medium through supernova explosions and calculate the 
average $^{60}$Fe mass yield from Type II supernovae (SNe II) to be 
about one-third of that for $^{26}$Al (\cite{tim95}; \cite{tim97}).  
This implies that, if the main contributor to the Galactic $^{26}$Al 
are supernovae, the 1173 and 1332 keV $^{60}$Fe line emissions should 
be close to the limits of detectability of current gamma-ray 
telescopes.  In this paper we present the observation of $^{26}$Al and 
$^{60}$Fe line emissions performed by the Gamma-Ray Imaging 
Spectrometer (GRIS).  While $^{26}$Al is clearly detected, we can only 
derive upper limits for the $^{60}$Fe.  We obtain the fluxes and 
Galactic abundance of these two isotopes assuming various source 
distribution models and we discuss the implications of the derived 
values on the current models of Galactic nucleosynthesis.

\section{The GRIS Flight}

\begin{figure}
\epsscale{.6}
\plotone{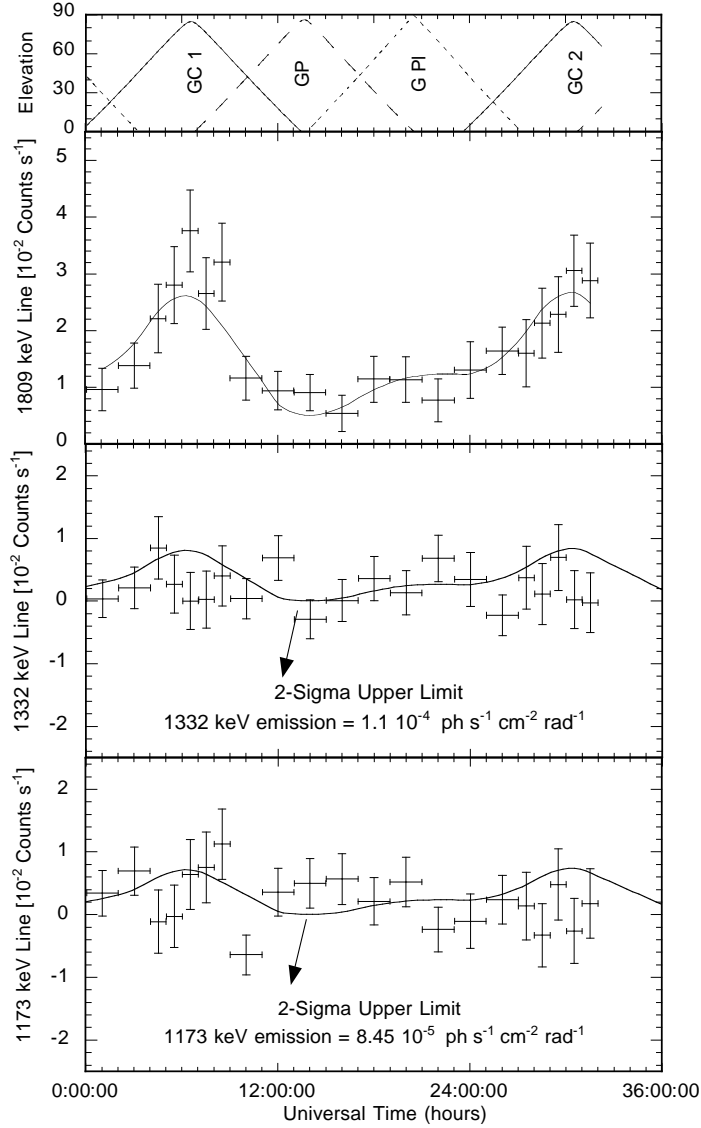}
\caption {Variation of the 1809, 1332 and 1173 keV 
line intensities during the flight.  These values include the 
instrumental background lines due to interactions of cosmic-ray 
induced neutrons with Al and Cu in the instrument.  Notice that while 
the 1809 keV line clearly shows a modulation due to the Galactic 
center transits, there is not a significant modulation for the 
$^{60}$Fe lines.  The solid line shows the predicted scan profile 
assuming a COS-B source model distribution.  The derived fluxes are 
$4.8\pm0.7\cdot10^{-4}$ photons s$^{-1}$ cm$^{-2}$ rad$^{-1}$ for the 
$^{26}$Fe emission and a combined 2-sigma upper limit of 
$6.9\cdot10^{-4}$ photons s$^{-1}$ cm$^{-2}$ rad$^{-1}$ for the 
$^{60}$Fe emission.\label{fig1}}
\end{figure}

The Gamma Ray Imaging Spectrometer (GRIS) is a balloon-borne 
high-resolution gamma-ray spectrometer consisting of an array of seven 
germanium detectors surrounded by a thick (15 cm) active NaI 
anticoincidence shield.  This instrument was reconfigured with a wide 
field collimator ($100^{\circ}\times 75^{\circ}$ FWHM field-of-view), 
and a 15 cm thick NaI blocking crystal to optimize its capability for 
observation of diffuse gamma-ray sources such as the cosmic diffuse 
background and Galactic line emissions, in particular the $^{26}$Al 
and the $^{60}$Fe radiations.  The measurements reported in this paper 
were made on a flight from Alice Springs, Australia, on 1995 October 
24-26.  The total germanium detector area and volume were 237 cm$^{2}$ 
and 1647.6 cm$^{3}$ respectively.  The duration at float altitude was 
32 hours at an average atmospheric depth of 3.8 g cm$^{-2}$.  The 
collimator was always pointed at the zenith.  From Alice Springs, the 
Galactic center, south Galactic pole, and Galactic plane 
(l=$240^{\circ}$) transit nearly overhead (see top of fig~\ref{fig1}).  
GRIS observed alternating 10 minute exposures with the blocking 
crystal open and closed.

\section{Data Analysis and Results}

The variation of the 1809, 1332 and 1173 keV line intensities measured 
during the flight is displayed in fig.  1.  These values were 
calculated by fitting the GRIS data with a Gaussian plus a power law 
model.  The intensity, centroid and width of the Gaussian were set as 
free parameters for the 1809 keV fits.  For the 1332 and 1173 keV 
fits, the centroids were fixed at the values expected from decay at 
rest and the widths were fixed at the values for narrow line emission.  
Since there was no hint of these lines in the spectrum, it was more 
appropriate to leave the line intensity as the only parameter of the 
fit.  The instrument energy resolution was precisely determined by 
fitting a line to the measured widths of many intense narrow lines in 
the background spectrum.  The derived instrumental width at 1809, 1332 
and 1173 keV were respectively 3.4, 2.8 and, 2.7 keV FWHM.

\placefigure{fig1}

Notice that the 1809 keV drift scan shows an excess during both 
Galactic center transits which is a detection of Galactic $^{26}$Al 
emission.  Such modulation is not observed in the 1332 and 1173 keV 
drift scan data.  Due to the lack of imaging capabilities of GRIS, the 
flux of the emissions was derived by fitting the drift scan data with 
a given source distribution model and taking into account the 
instrument response plus the effects of atmospheric absorption.  Based 
on recent nucleosynthesis and stellar evolution models (\cite{tim97}) 
we have assumed that most of the Galactic $^{26}$Al and $^{60}$Fe is 
generated by SNII explosions and thus, both isotopes have an identical 
spatial distribution.  This assumption is also supported by the 
COMPTEL irregular profile that favors massive stars as the source of 
$^{26}$Al.  In order to quantify the influence of the source 
distribution profile on the derived fluxes, we have extended the study 
to several models.

Fig ~\ref{fig2} shows the latitude integrated flux profile for the 
models considered.  The long-dashed line corresponds to the Galactic 
high-energy-gamma-ray measurement performed by the COS-B instrument, 
which has been used to fit most previous 1809 keV observations.  The 
dotted line represents the profile derived from the 1.8 MeV COMPTEL 
map from 3.5 yr of observation integrated for 
$-10^{\circ}<b<10^{\circ}$ (\cite{obe96}).  The solid line 
corresponds to the free electron distribution of Taylor \& Cordes 
(1993), that shows the spiral structures and has previously been used 
for explaining some of the features on the $^{26}$Al COMPTEL map with 
the addition of a Galactic bar component (\cite{chen96}).  The 
short-dashed curve corresponds to an exponential distribution with 4.5 
kpc scale radius, which is a good representation of the stellar disk 
and is well fit to the COMPTEL map (\cite {die95}).  Notice that the 
Galactic abundances are a byproduct of the calculation when using 
3-dimensional models.  The fluctuations outside the Galactic center 
region shown by the exponential model are due to local isolated SN 
events.  The intensity and location of theses are specific to the 
random sequence used for the generation of the galaxy model.  The 
peaks shown in this particular distribution represent about 20\%
of the emission from the central radian which gives an idea of the 
contribution that local sources could have on the total emission.

\placefigure{fig2}

The fluxes and abundances resulting from the different models are 
displayed in table ~\ref{tbl-2}.  The 1809 keV line flux derived from 
the COS-B distribution (as used in \cite{ma84}) is 
$4.8\pm0.7\cdot10^{-4}$ photons s$^{-1}$ cm$^{-2}$ rad$^{-1}$.  This 
value is consistent with the fluxes reported by previous observations 
and it constitutes the most statistically significant flux measurement 
performed with high energy resolution (6.8$\sigma$ confidence level).  
On the other hand, the 1332 and 1173 keV best fit fluxes are 
$1.6\pm4.8\cdot10^{-5}$ and $-2.5\pm5.5\cdot10^{-5}$ photons s$^{-1}$ 
cm$^{-2}$ rad$^{-1}$ which are compatible with no detection of 
Galactic $^{60}$Fe.  The derived 2$\sigma$ flux upper limits for 1332 
keV and 1173 keV line emission are respectively $1.1\cdot10^{-4}$ and 
$8.5\cdot10^{-5}$ photons s$^{-1}$ cm$^{-2}$ rad$^{-1}$.  Since these 
two photons are emitted simultaneously in every $^{60}$Fe decay, we 
derive a combined 2$\sigma$ upper limit of $6.8\cdot10^{-5}$ photons 
s$^{-1}$ cm$^{-2}$ rad$^{-1}$.  The fluxes obtained from the other 
models are similar to those derived from the COS-B distribution.  
Making a joint fit to the $^{26}$Al and $^{60}$Fe drift scan we derive 
a $^{60}$Fe to $^{26}$Al flux ratio 2$\sigma$ upper limit of 0.14.

The derived fluxes and upper limits are the most significant ever 
measured by a high resolution gamma-ray spectrometer and are similar 
to those reported by the satellite instruments SMM (\cite{lei94}) and 
OSSE (\cite{har97}).  However, we must point out that, unlike in 
previous observations, the error on this measurement is dominated by 
statistical uncertainties.  The systematic uncertainties are small 
because the data comes from a single balloon flight, with low 
background and a high signal to background ratio.  Furthermore, the 
count rate spectra do not show any significant background lines at the 
energies of interest and, the continuum background was very stable 
during the whole flight.

\placetable{tbl-2}

The measured astrophysical spectrum around the 1809, 1332 and 1173 keV 
lines is displayed in fig~\ref{fig3}.  This spectrum was derived by 
subtracting the Galactic pole and Galactic plane accumulation from the 
sum of both Galactic center transits.  Since we do not expect 
significant line emission from the Galactic pole and the Galactic 
plane (l=$240^{\circ}$) the subtracted spectrum should be mostly of 
Galactic origin.  The different energy regions were fit with a 
Gaussian profile plus a constant (see the resulting fit parameters 
included in the figure).  The 1809 keV line is well resolved and the 
details of the analysis and the implications of this detection can be 
found in a recent work by Naya \etal\  (1996).  The most remarkable 
result is the intrinsic width of the line (5.4 [+1.4,-1.3] keV FWHM) 
which is approximately three times broader than expected from the 
effect of Doppler broadening due to Galactic rotation (\cite{ski91}, 
\cite{ge96}).  This large width implies that the $^{26}$Al is moving 
at velocities of $>$450 km s$^{-1}$ and it favors models of an origin 
of the $^{26}$Al in Supernovae or Wolf-Rayet stars rather than from 
the slower winds in AGB stars.  However, it is not well understood how 
$^{26}$Al can maintain such a high speed for 10$^{6}$ years.  
Different scenarios that can account for a broad emission have been 
studied, but none of them seems to provide a satisfactory explanation 
to the GRIS observation (\cite{chen97}).  For the $^{60}$Fe lines 
(fig~\ref{fig3} (b) and (c)) only the line intensity was left as a free 
parameter.  The best fit is consistent with no $^{60}$Fe detection, 
which is in good agreement with the drift scan analysis previously 
shown.

\placefigure{fig3}

\section{Discussion}

\begin{figure}
\epsscale{.7} \plotone{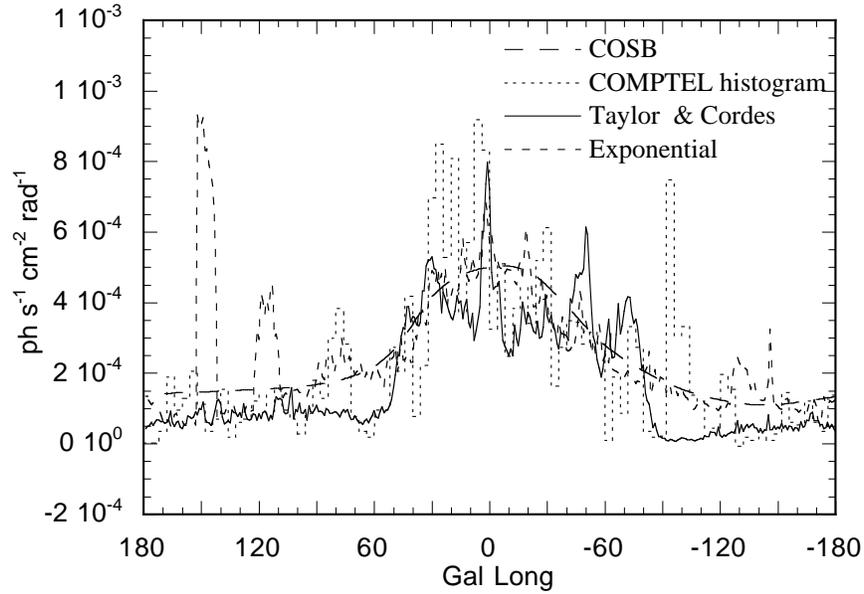} \caption {Longitude flux 
distributions considered for the flux and abundance studies presented 
herein.  The curves are normalized to the value that fits the GRIS 
1809 keV drift scan data.  The long-dashed line is based on the 
high-energy gamma ray measurement performed by the COS-B instrument 
(\cite{may82}).  The dotted line corresponds to the longitude profile 
derived from the COMPTEL map (\cite{obe96}).  The solid and 
short-dashed lines have been derived from the sum of SN events 
generated by Monte-Carlo technique following 3-dimensional models that 
are a reasonable fit to the COMPTEL map such as the Taylor \& Cordes 
distribution (1993) and an exponential distribution with 4.5 kpc scale 
radius respectively.
\label{fig2}}
\end{figure}

\begin{figure}
\epsscale{.5}
\plotone{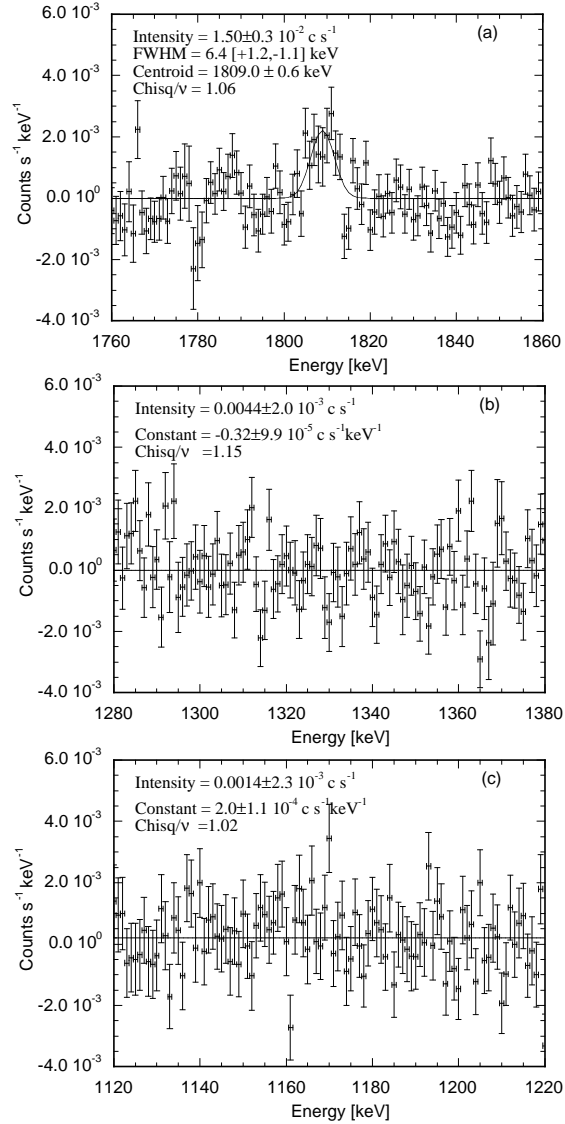}
\caption {Net Galactic Center count rate spectrum 
around the 1809, 1332 and 1173 keV energies.  Notice the clear 
detection of the Galactic $^{26}$Al emission at 1809 keV and the absence of 
a significant excess for the $^{60}$Fe lines.  The solid curve is the best 
fit of the data to a Gaussian line shape.  The derived intrinsic width 
for the astrophysical 1809 keV line is 5.4 [+1.4,-1.3] keV FWHM, which 
is more than three times the value expected from previous theories 
(see a more detailed discussion in Naya \etal\  1996 ).\label{fig3}}
\end{figure}

The derived $^{26}$Al and $^{60}$Fe abundances and fluxes shown in 
table ~\ref{tbl-2} represent a valuable piece of information for the 
understanding of the nucleosynthetic activity in our Galaxy.  Notice 
however, that these values are closely related to the assumed models 
which, in principle, could differ significantly from reality.  The 
obtained $^{26}$Al abundances are significantly higher than the 1.8 
$\, M_\odot$ reported by COMPTEL (\cite{kno98}).  This discrepancy 
results from the $^{26}$Al emission flux derived by GRIS, which is 
higher than that derived by COMPTEL. Taking the measured COMPTEL 
longitude distribution as the model the source distribution (see 
fig~\ref{fig2}) we derive a flux which is $5.4\pm0.7\cdot10^{-4}$ photons 
s$^{-1}$ cm$^{-2}$ rad$^{-1}$ which is 2.8 times higher than that 
derived for the inner Galaxy l=[328,35] and b=[-5,5] in the COMPTEL 
map (\cite{obe96}).  Furthermore, a comparison with the fluxes 
measured by all other instruments shows that COMPTEL has the lowest 
flux value reported (\cite{pra96}).  This suggests that a significant 
fraction of the $^{26}$Al emission is not shown in the COMPTEL map.  
This is not surprising since COMPTEL has little sensitivity to low 
surface brightness extended emission.  In a recent study it has been 
shown that the existence of dispersed $^{26}$Al from nearby SN or 
$^{26}$Al confined to fragments located at medium latitude that do not 
appear clearly in the map cannot be ruled out with the COMPTEL data 
(\cite{kno97}).  A combined analysis of the GRIS and COMPTEL 
observations is currently being performed in order to refine the 
comparison of GRIS and COMPTEL results.

We have studied the effect that known local sources could have on the 
results.  The Cygnus Loop, one of the local sources identified in the 
COMPTEL map, should not contribute to the observed emission since this 
source was well outside the GRIS FOV during the entire flight.  Vela, 
the other local source identified in the COMPTEL map, is within the 
GRIS FOV during the Galactic plane transit but, is so weak that it has 
a negligible contribution to the measured 1809 keV count rate.  
Another potential local source is Loop I in the Sco-Cen association, 
which has an angular diameter of 116° centered at 170 pc from the Sun 
and has been suggested to be the source of 1809 keV emission 
(\cite{bd89}).  Recent ROSAT and radio observations suggest a mean 
rate of 2 SNIIs per 10$^{6}$ y in the Sco-Cen association and indicate 
that the latest SNII may have occurred $\approx 2\cdot10^{5}$ y ago 
with a progenitor mass of 15-20 $\, M_\odot$ (\cite{egg93}).  Assuming 
that these events ejected about 10$^{-4}$ $\, M_\odot$ of $^{26}$Al, 
the resulting emission would be more than one order of magnitude too 
weak to account for the 1809 keV line rates observed by GRIS. Another 
argument for discarding Loop I as an important source of the observed 
1809 keV line is based on the drift scan profile it would produce.  
The maximum emission from Loop I should be shifted 3 hours with 
respect to the maximum emission expected from a source located in the 
Galactic center region.  We have calculated the maximum contribution 
from Loop I that is compatible with the data.  For this calculation we 
fit the drift scan data with a model based on the COS-B distribution 
plus Loop I, leaving the intensity of both components as free 
parameters.  The resulting best fit is compatible with no contribution 
from Loop I and the contribution to the total intensity should be 
under 20\% at the 2$\sigma$ confidence level.

Another factor that could alter the results is the width of the $^{60}$Fe 
emission.  We have shown that the $^{26}$Al line was detected with an 
intrinsic width of 0.3\% FWHM. For the present analysis, we have 
assumed the $^{60}$Fe emission to be narrow but, in principle, it may 
present a broadening like the $^{26}$Al line does.  It is reasonable to 
assume that any broadening should not be more than 0.3\%, since $^{60}$Fe 
has a longer lifetime than $^{26}$Al and therefore has more time to slow 
down before decaying.  Such a broadening would make the continuum 
background under the lines to be 1.7 times more intense than for the 
corresponding in a narrow line.  The derived upper limits would be 
therefore $\sqrt{1.7}=1.3$ times higher than those presented in this 
work.

The comparison of the measured fluxes with those predicted by 
theoretical models (\cite{tim97}) has interesting implications.  The 
measured 2$\sigma$ upper limit for the $^{60}$Fe to $^{26}$Al flux 
ratio is 0.14 and is relatively independent of the assumed source 
distribution.  This ratio is slightly lower than the 0.16 ratio 
predicted in a recent work by Timmes \& Woosley (1997).  The 
predicted $^{60}$Fe and $^{26}$Al fluxes were calculated by averaging 
the mass of ejected $^{60}$Fe and $^{26}$Al from Woosley \& Weaver 
(1995) star models over a Salpeter initial mass function 
(\cite{tim97}).  The flux uncertainties claimed for these calculations 
are up to a factor of 2.  These uncertainties induce an error on the 
predicted ratio of $\pm0.1$ (\cite{die97}) which makes our measured 
upper limit to be compatible with the expectations.  On the other 
hand, Meynet \etal\ (1997) have recently estimated that 20\%-70\% of
the Galactic $^{26}$Al could be
made in the hydrogen envelope of
Wolf-Rayet stars.  Since these sources do not release significant 
amounts of $^{60}$Fe, they could well account for a lower $^{60}$Fe to 
$^{26}$Al flux ratio.

The measurement of the $^{60}$Fe emission remains a crucial objective for 
future generations of gamma-ray telescopes.  Its detection would be 
very valuable to determine the contribution of the different sources 
to the production of $^{26}$Al, which would provide the most accurate 
information about the current nucleosynthetic activity in our Galaxy.  
Instruments such as the spectrometer of the ESAÕs INTEGRAL project 
(\cite{pvb95}), the High Energy Solar Spectroscopic Imager 
(\cite{den96}) and, the Long Duration Balloon GRIS (Marvin 
Leventhal private communication) should be able to address this issue 
at the beginning of the next decade.

\acknowledgments

Our thanks to the GRIS experiment development team, Stephen Deredyn, 
Stephen Snodgrass, Chris Miller and Kiran Patel.  Launch and recovery 
support was provided by the crew of the National Scientific Ballooning 
Facility. We also thank an anonymous referee for his helpful comments.

\clearpage
 
\begin{table*}
\caption{ $^{26}$Al and $^{60}$Fe Galactic Flux and Abundance Derived From GRIS. \label{tbl-2}}
\begin{center} 
\begin{tabular}{lccccc}
\hline \hline
\multicolumn{1}{c|}{} &\multicolumn{2}{c|}{Flux} &\multicolumn{1}{c|}{$^{60}$Fe/$^{26}$Al} 
&\multicolumn{2}{c}{Abundance}\\
\multicolumn{1}{c|}{} &\multicolumn{2}{c|}{$10^{-4}$ ph s$^{-1}$cm$^{-2}$rad$^{-1}$} 
&\multicolumn{1}{c|}{Flux Ratio} 
&\multicolumn{2}{c}{$M_\odot$}\\
\hline
\multicolumn{1}{l|}{Model} &\multicolumn{1}{c}{$^{26}$Al} &\multicolumn{1}{c|}{$^{60}$Fe
\tablenotemark{a}} 
&\multicolumn{1}{c|}{} &\multicolumn{1}{c}{$^{26}$Al} &\multicolumn{1}{c}{$^{60}$Fe\tablenotemark{a}}\\
\hline
COS-B\tablenotemark{b}	       & $4.87\pm0.72$	& $<$0.68   & $<$0.14   & -	            & - \\
COMPTEL\tablenotemark{b}       & $5.48\pm0.78$	& $<$0.72   & $<$0.13   & -             & - \\ 
T\&C	                       & $3.97\pm0.59$	& $<$0.54   & $<$0.14   & $2.61\pm0.39$	& $<$1.69 \\
Exponential	                   & $4.97\pm0.74$	& $<$0.71   & $<$0.14   & $4.52\pm0.67$	& $<$3.11 \\
\hline \hline
\end{tabular}
\end{center}
\tablenotetext{a}{2$\sigma$ upper limits}
\tablenotetext{b}{Abundances can not be derived from 2-dimensional 
models}
\tablenum{1A}
\end{table*}

\clearpage

\clearpage

\clearpage

\end{document}